\documentclass[useAMS,usenatbib]{mn2e}
\usepackage{graphicx,color,epsfig}
\usepackage{color}

\def\ltsima{$\; \buildrel < \over \sim \;$}
\def\simlt{\lower.5ex\hbox{\ltsima}}
\def\gtsima{$\; \buildrel > \over \sim \;$}
\def\simgt{\lower.5ex\hbox{\gtsima}}

\title[The Accretion Disc Particle Method]
      {The Accretion Disc Particle Method for Simulations of Black Hole Feeding and Feedback}
      
      \author[C. Power, S. Nayakshin \& A. R. King]
             {Chris Power\thanks{chris.power@astro.le.ac.uk}, 
                 Sergei Nayakshin \& Andrew King\\ 
                 Department of Physics \& Astronomy,
                 University of Leicester, Leicester, LE1 7RH, UK}

\begin{document}
      
\date{}
      
\pagerange{\pageref{firstpage}--\pageref{lastpage}} \pubyear{2010}
      
\maketitle
      
\label{firstpage}

\begin{abstract} 

Black holes grow by accreting matter from their surroundings. However, 
angular momentum provides an efficient natural barrier to accretion and so 
only the lowest angular momentum material will be available to feed the 
black holes. The standard sub-grid model for black hole accretion in galaxy 
formation simulations -- based on the Bondi-Hoyle method -- does not account 
for the angular momentum of accreting material, and so it is not clear how 
representative the black hole accretion rate estimated in this way is likely to
be. In this paper we introduce a new sub-grid model for black hole accretion 
that naturally accounts for the angular momentum of accreting material. Both 
the black hole and its accretion disc are modelled as a composite 
\emph{accretion disc particle}. Gas particles are captured by the 
accretion disc particle if and only if their orbits bring them within its 
accretion radius $R_{\rm acc}$, at which point their mass is added to the 
accretion disc and feeds the black hole on a viscous timescale 
$t_{\rm visc}$. The resulting black hole accretion rate $\dot{M}_{\rm BH}$ 
powers the accretion luminosity $L_{\rm acc} \propto \dot{M}_{\rm BH}$, which 
drives black hole feedback. Using a series of controlled numerical 
experiments, we demonstrate that our new accretion disc particle method is 
more physically self-consistent than the Bondi-Hoyle method. We also discuss 
the physical implications of the accretion disc particle method for systems 
with a high degree of rotational support, and we argue that the 
$M_{\rm BH}-\sigma$ relation in these systems should be offset from the relation
for classical bulges and ellipticals, as appears to be observed.
\end{abstract}


\section{Introduction}
\label{sec:intro}

Understanding how super-massive black holes at the centres of galaxies 
grow over cosmic time is one of the most important yet challenging problems 
facing modellers of galaxy formation. Observationally there is clear and 
compelling  evidence that in galaxies that host super-massive black 
holes the black hole mass $M_{\rm BH}$ correlates strongly with the stellar 
mass $M_{\ast}$ and velocity dispersion $\sigma$ of the host bulge 
\citep[e.g.][]{magorrian.1998,ferrarese.2000,gebhardt.2000,tremaine.2002,
haring.rix.2004,gultekin.2009}. Theoretically these correlations are widely 
interpreted as hallmarks of black hole feedback, which itself is a natural 
consequence of accretion onto the black hole \citep[e.g.][]{silk.rees.1998,
fabian.1999,king.2003,sazonov.etal.2005,king.2005}. In this picture, feedback 
acts to regulate the black hole's mass accretion rate $\dot M_{\rm BH}$ by 
modifying the physical and dynamical state of gas in and around its host 
galaxy -- so the greater $\dot M_{\rm BH}$, the stronger the feedback and the 
greater the impact on $\dot M_{\rm BH}$. Therefore, how one estimates 
$\dot M_{\rm BH}$ is crucial because it governs not only the rate at which the 
black hole grows but also the strength of the black hole feedback. This is a 
particularly important problem because how black hole feeding and feedback is 
modelled can have a profound impact on the predictions of how galaxies form 
\citep[e.g.][]{bower.etal.2006,croton.etal.2006}.\\

The standard approach to estimating $\dot M_{\rm BH}$ in galaxy formation
simulations is based on the work of \citet{bondi.hoyle.1944}
and \citet{bondi.1952} \citep[hereafter the Bondi-Hoyle method; 
  cf.][]{dimatteo.2005,springel.etal.2005}. In the accretion problem as it was
originally formulated, a spherically symmetric accretion flow is captured
gravitationally by a point-like accretor from a 
uniform distribution of gas with zero angular momentum. Under these conditions,
the accretion rate onto the accretor $\dot M_{\rm Bondi}$ is proportional to the 
square of the black hole mass $M_{\rm BH}^2$ and the gas density $\rho$, and 
inversely proportional to the cube of the sound speed $c_s$. This gives an
accretion rate $\dot{M}_{\rm Bondi} \propto M_{\rm BH}^2 \rho/c_s^3$. The 
assumption in galaxy formation simulations is that $\dot M_{\rm BH} \propto 
\dot M_{\rm Bondi}$ \citep[see, for example, the discussion 
in][]{booth.schaye.2009}.

However, there are good physical reasons to believe that $\dot M_{\rm Bondi}$ 
cannot be representative of the true black hole accretion rate $\dot M_{\rm BH}$
in an astrophysically realistic situation \citep[cf.][]{king.2010}. First, 
the black hole is embedded in the gravitational potential of a galaxy that is 
orders of magnitude more massive than it; this means that the gravitational 
force acting on the accretion flow is dominated by the mass of the galaxy 
rather than the black hole and so $\dot M_{\rm Bondi}$ will be a similar number 
of orders of magnitude off the true $\dot M_{\rm BH}$ (we show this explicitly 
in Hobbs et al., in preparation). Second, any astrophysically realistic 
accretion flow will have some angular momentum, violating one of the key
assumptions made in calculating $\dot M_{\rm Bondi}$. This is important
because it
implies that infalling material will settle onto a circular orbit whose radius 
$R_{\rm circ}$ is set by the angular momentum of the material with respect to 
the black hole \citep[cf.][]{hobbs.etal.2010a}. In particular, it means that
only the very lowest angular material will be available to feed the black hole
because the timescale required for viscous transport of material through
the disc is of order a Hubble time on scales of order $R\sim 1-10\rm pc$ 
\citep[see, for example, ][]{king.2010}. This is a very restrictive condition 
because it is not straightforward for infalling
gas to lose its angular momentum other than by colliding with other gas, which 
leads to angular momentum cancellation. Therefore, angular momentum provides 
an efficient natural barrier to accretion by the black hole, and so must be 
accounted for when estimating $\dot M_{\rm BH}$.\\

These arguments make clear that the Bondi-Hoyle method cannot provide a reliable
estimate of $\dot M_{\rm BH}$ in galaxy formation simulations. If feedback from
black holes plays as important a role in galaxy formation as we expect it to
\citep[e.g.][]{bower.etal.2006,croton.etal.2006}, then it is crucial that
we devise an alternative method for estimating $\dot M_{\rm BH}$ in galaxy 
formation simulations that overcomes the problems that beset the Bondi-Hoyle 
method.

In this short paper, we introduce our new ``accretion disc particle'' method 
(hereafter the ADP method) for estimating $\dot M_{\rm BH}$ in galaxy formation 
simulations, which accounts naturally for the angular momentum of infalling 
material. We use a collisionless accretion disc particle (ADP) to model the 
black hole and its accretion disc. The black hole accretes if and only if gas 
comes within the accretion radius $R_{\rm acc}$ of the ADP, at which point 
it is captured and added to the accretion disc that feeds the black hole on a 
viscous timescale $t_{\rm visc}$. In this way the black hole will accrete only 
the lowest angular momentum material from its surroundings.

The layout of this paper is as follows. We describe the main features of the 
ADP method in \S\ref{sec:accretion_models}, showing how the accretion rate 
$\dot M_{\rm acc}$ onto the ADP is linked to the black hole accretion rate 
$\dot M_{\rm BH}$. In \S\ref{sec:feedback} we discuss briefly our 
momentum-driven feedback model \citep[cf.][]{nayakshin.power.2010} as well 
as our implementation of the 
quasar pre-heating model of \citet{sazonov.etal.2005}. The accretion rate 
$\dot M_{\rm BH}$ estimated using the ADP method is very different from one 
estimated using the Bondi-Hoyle method. We show this clearly in 
\S\ref{sec:results} using simple idealised numerical simulations, designed 
to illustrate the key differences between the ADP and Bondi-Hoyle methods for 
estimating 
$\dot M_{\rm BH}$. These simulations follow the collapse of an 
initially rotating shell of gas onto a black hole embedded in an isothermal 
galactic potential. Finally we summarise our results in \S\ref{sec:summary} 
and we discuss the implications for galaxy formation simulations and 
the $M_{\rm BH}-\sigma$ relation in \S\ref{sec:conclusions}.

\section{Modelling Accretion and Feedback}
\label{sec:accretion_models}

\subsection{The Accretion Model}
\label{ssec:accretion}

\subsubsection{The Accretion Disc Particle (ADP) Method}

\begin{figure}
\centerline{\psfig{file=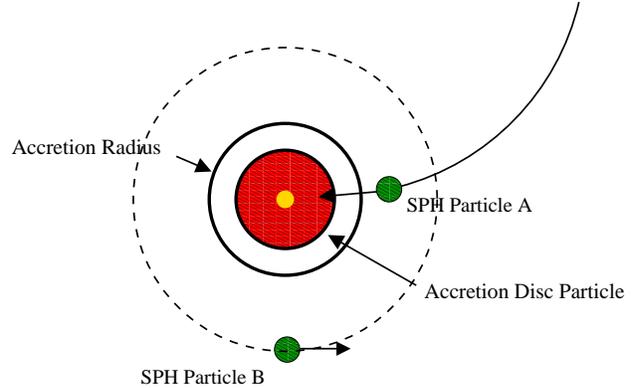,width=0.47\textwidth,angle=0}}
\caption{{\bf The Accretion Disc Particle (ADP) Method.} The ADP is a 
  collisionless sink particle that consists of a black hole and its 
  accretion disc. SPH particle A has a small angular momentum and so its 
  orbit brings it within the ADP's accretion radius $R_{\rm acc}$, at which 
  point it is added to the accretion disc. SPH particle B's angular momentum 
  is too large for it to be captured. The black hole feeds from the accretion 
  disc on a viscous timescale $t_{\rm visc}$. Both $R_{\rm acc}$ and $t_{\rm visc}$ 
  are free parameters in the accretion disc particle method. See text for 
  further details.
}
\label{fig:accretion_model}
\end{figure}

The main features of the ADP are illustrated in Fig~\ref{fig:accretion_model}. 
The ADP is collisionless and consists of a sink particle \citep{bate.1995} 
with an accretion radius $R_{\rm acc}$. $R_{\rm acc}$ is a free parameter of the 
simulation but in general it is desirable to set it to the smallest resolvable 
scale in the simulation, which will be of order the gravitational softening 
length of the gas particles. The total mass of the sink particle is equal to 
the sum of the masses of the black hole, $M_{\rm BH}$, and its accretion disc, 
$M_{\rm disc}$. The accretion disc is assumed to be tightly bound to the black 
hole and is thus a property of the sink particle rather than a separate entity. 

Accretion onto the black hole in the ADP method is a two-stage process. First,
any gas that crosses the accretion radius $R_{\rm acc}$ is removed from the 
computational domain and added to the accretion disc. In the classical 
sink-particle method of \citet{bate.1995}, the accreted gas would be added 
to the black hole immediately, but in an astrophysically realistic situation, 
the finite non-zero angular momentum of the accreted gas leads to the 
formation of an accretion disc before the gas can accrete onto the black hole. 
Here we assume that gas is added to the accretion disc after a time that is
of order the dynamical timescale $t_{\rm dyn}$ at $R_{\rm acc}$.

Second, gas is transported through the accretion disc and is added to the black
hole. In principle, we could describe the evolution of the accretion disc by 
the standard viscous disc evolution equations \citep[see, for example, 
Chapter 5 of][]{frank.king.raine}. However, neither theory nor observation
tell us what the magnitude of the disc viscosity should be and so we cannot 
be sure of the efficiency of angular momentum transport within the disc. 
Moreover, if the accretion disc is sufficiently massive to become 
self-gravitating, then a significant fraction of its gas mass can be converted 
to stars \citep[see, for 
example,][]{toomre.1964,paczynski.1978,shlosman.1989,goodman.2003,
nayakshin.etal.2007}. Star formation depletes the gas reservoir, reducing 
$\dot{M}_{\rm BH}$, but stellar feedback can act to either reduce or enhance 
$\dot{M}_{\rm BH}$ \citep[e.g.][]{cuadra.etal.2008,schartmann.etal.2009}. 
Solving the standard viscous disc evolution equations is no longer 
straightforward under these circumstances and detailed accretion disc 
simulations are required \citep[see, for example ][]{cuadra.etal.2008}. 

We want a simple sub-grid model, however, and so we simplify the problem.
To this end, we assume that (a) an accretion disc forms and (b) angular 
momentum transport through the disc introduces a delay between the time a 
gas particle crosses $R_{\rm acc}$ and the time that it is 
accreted by the black hole. This time delay will be of the order of the disc 
viscous time, $t_{\rm visc}$, which can be of order the Hubble time for accretion
discs around super-massive black holes \citep[cf.][]{king.2010}. To capture 
this in the simplest way we describe the evolution of the accretion disc by
\begin{equation}
 \label{eq:dmdiscdt}
 \dot{M}_{\rm disc} = \dot{M}_{\rm acc} - \dot{M}_{\rm BH}\;,
\end{equation}
\noindent where $\dot M_{\rm acc}$ is the rate at which gas is captured
through $R_{\rm acc}$ and $\dot M_{\rm BH}$ is the accretion rate onto the
black hole. $t_{\rm visc}$ can be estimated using physical arguments 
\citep[see, e.g. \S2 of][]{king.2010} but, as argued above, it is uncertain, 
and so we treat $t_{\rm visc}$ as a free parameter instead. In general we 
require that $t_{\rm visc} \simgt t_{\rm dyn}(R_{\rm acc})$ and typically set
$t_{\rm visc} \sim 10-100 \rm Myrs$ in galaxy formation simulations 
(Power et al., in preparation).

Note that the rate at which gas is captured from the ambient medium is not 
limited in any way; it is simply governed by the evolution of the large
scale accretion flow into the centre of the galaxy. In contrast, the rate at 
which gas is accreted onto the black hole is limited by the Eddington 
accretion rate $\dot M_{\rm Edd}$,
\begin{equation}
 \label{eq:eddington}
 \dot{M}_{\rm Edd} \, \equiv \, {{4\pi \, G \, M_{\rm BH} \, m_{\rm p}}
\over {\eta \, \sigma_{\rm T} \, c}} \, ,
\end{equation}
\noindent where $m_{\rm p}$ is the proton mass, $\sigma_{\rm T}$ is the
Thomson cross-section, $c$ is the speed of light and $\eta$ is the 
accretion efficiency, for which we assume the standard value of $\eta=0.1$. 
This means that $\dot M_{\rm BH}$ in equation~\ref{eq:dmdiscdt} satisfies
\begin{equation}
  \label{eq:dmbhdt}
  \dot M_{\rm BH} = \min\left[ \frac{M_{\rm disc}}{t_{\rm visc}}\;,
    \dot M_{\rm Edd} \right]
\end{equation}
\noindent This simple system of equations can be expanded in the future to
encompass more detailed disc modelling, including gas disc self-gravity and the
resulting star formation and feedback from stars formed there. Given the
empirical evidence from our Galactic Centre 
\citep[e.g.,][]{nayakshin.2005,paumard.2006} and the theoretical expectation 
that nuclear stellar cluster feedback should be important 
\citep[cf.][]{comp.feedback.paper}, this is likely to be an important step in 
future studies.\\

It is worth making some additional comments about our ADP method
and how it relates to the classical sink particle formulation of 
\cite{bate.1995}. In the classical sink particle formulation, a number 
of conditions had to be satisfied before gas could be accreted by the sink 
particle (e.g. pressure forces at the accretion radius, comparison of thermal 
and gravitational binding energy with respect to the sink particle, etc...). 
In our approach, there is a single condition for accretion, namely that gas
comes within $R_{\rm acc}$. Physically this is quite reasonable. The 
virial temperature in the vicinity of a super-massive black hole is very
high, typically in the range of $10^6 - 10^8$ K. However, gas densities are 
also very high near $R_{\rm acc}$ and so cooling times are expected to be very 
short \citep[e.g.][]{king.2005}. This implies that gas is likely to
be much cooler than the virial temperature, which means that both the
pressure forces and thermal energy of the gas is negligible. Furthermore,
viscous times are always very long compared with dynamical times, so we 
expect the accretion disc to be a long-lived (essentially permanent) feature 
within $R_{\rm acc}$. This means that gas that comes within $R_{\rm acc}$
is very likely to undergo a large Mach number collision with the disc, 
causing it to shock and then cool rapidly. Therefore, even if gas is
initially on an unbound (hyperbolic) orbit around the sink particle, it will
most likely lose most of its bulk and thermal energy and settle into the
accretion disc. 

\subsubsection{The Bondi-Hoyle Method}

The Bondi-Hoyle method for estimating $\dot M_{\rm BH}$ is the standard approach
in galaxy formation simulations \citep[see, for example, ][]{dimatteo.2005,springel.2005,booth.schaye.2009}. Here the black hole accretion rate is calculated 
directly from
\begin{equation}
  \label{eq:bondi_hoyle}
  \dot{M}_{\rm BH} \, = \, {{4\pi \, \alpha \, G^2 M_{\rm BH}^2 \, \rho} 
    \over {(c_s^2 + v^2)^{3/2}}} \, ,
\end{equation}
\noindent where $\rho$ is the SPH density at the position of the black hole,
$c_s$ is the sound speed of the gas, $v$ is the velocity of the black hole 
relative to the gas and $\alpha$ is a fudge factor that we set to unity for 
the purposes of this work, but which can be of order $\sim 100-300$ 
\citep[see the discussion in][]{booth.schaye.2009}. 
In practice we compute estimates for $\rho$, $c_s$ and $v$ using the SPH 
smoothing kernel with $N_{\rm SPH}$=40 neighbours. Note that there is no 
explicit dependence on the angular momentum of the gas in 
equation~\ref{eq:bondi_hoyle} -- the accretion rate is dictated by the 
gas density $\rho$ and sound speed $c_s$.

\subsection{The Feedback Model}
\label{sec:feedback}

In the simulations presented in the next section, we use $\dot M_{\rm BH}$ 
estimated using either equation~\ref{eq:dmbhdt} or \ref{eq:bondi_hoyle} 
to determine the accretion luminosity of the black hole,

\begin{equation}
\label{eq:luminosity}
L_{\rm acc} = \eta \dot{M}_{\rm BH} c^2;
\end{equation}

\noindent this is Eddington limited, as explained in the previous section. We 
assume that this radiated luminosity drives a wind that carries a momentum flux
$L_{\rm acc}/c$, which is usually true for AGN 
\citep[cf.][]{king.pounds.2003,king.2010}. Wind particles are emitted 
isotropically by the black hole at a rate
\begin{equation}
\dot N_{\rm wind} = \frac{L_{\rm acc}}{c p_{\rm wind}}\;,
\label{nrate}
\end{equation}
\noindent and they carry a momentum $p_{\rm wind} = 0.1 m_{\rm gas} \sigma$, 
where $\sigma$ is the velocity dispersion of the host halo. This satisfies 
the requirement $p_{\rm wind} \ll p_{\rm gas}$, where 
$p_{\rm gas}$ is the typical gas particle momentum ($\sim m_{\rm gas} 
\sigma$ here), and ensures that Poisson noise from our Monte Carlo scheme 
does not compromise our results \citep[see][]{rad.trans.paper}.

In addition to this momentum-driven wind, we include the quasar pre-heating
model of \citet{sazonov.etal.2005}. In this model, the average 
quasar spectral energy distribution derived by \citet{sazonov.etal.2004} is
used to estimate an equilibrium temperature $T_{\rm eq}$ for the gas
based on the ionisation parameter $\xi(r)=L_{\rm acc}/n(r)r^2$, where $n(r)$ 
is the number density at radius $r$. Physically $T_{\rm eq}$ corresponds to 
the temperature at which heating through Compton scattering and photoionisation
balances Compton cooling and cooling as a result of continuum and line 
emission, on the assumption that gas is in ionisation equilibrium. In practice,
we calculate heating and cooling rates using formulae A32 to A39 in Appendix 
3.3 of \citet{sazonov.etal.2005}, and we find that the resulting equilibrium 
temperature profile of the gas is well approximated by their equation 3,
\begin{equation}
  \label{eq:temp_eq}
  {T_{\rm eq}(\xi) \simeq 200 \xi \rm K.}
\end{equation}
\noindent This holds over the temperature range $2\times10^4\rm K$ to 
$10^7\rm K$; for $\xi\ll 100$ and $\xi \gg 5 \times 10^4$, $T_{\rm eq} 
\simeq 10^4\rm K$ and $2\times 10^7\rm K$ respectively.  

\section{Results}
\label{sec:results}

We have run simple idealised numerical simulations that are designed to
show that our ADP method constitutes a physically self-consistent sub-grid
model for estimating the black hole accretion rate $\dot M_{\rm BH}$ in
galaxy formation simulations, and that the Bondi-Hoyle method does not.

Our initial condition is a spherical shell of 
gas of a uniform density $\rho_0$, distributed between the inner and outer 
radii, $R_{\rm in}$ and $R_{\rm out}$, respectively. The shell is embedded in the 
static analytic gravitational potential of a singular isothermal sphere with a 
1-D velocity dispersion $\sigma$ and modified slightly to have a constant 
density core within $R \le R_{\rm core}$. For 
all the runs in this paper, we adopt $\rho_0 \simeq 10^{10} \rm
M_{\odot}\,kpc^{-3}$, $R_{ \rm in} = 0.067$ kpc, $R_{\rm out}=0.1$ kpc, 
$R_{\rm core}=0.01$ kpc and $\sigma=147 \rm kms^{-1}$. The shell has a mass of 
$M_{\rm shell}=3\times 10^7\rm M_{\odot}$ and is realised with $\sim 280,000$ gas
particles, drawn from a uniform density glass, which means that the particle 
mass is $m_{\rm gas} \simeq 1.1 \times 10^2 \,M_{\odot}$. We give the shell an 
initial temperature of $10^4$ K and an initial bulk rotation around the 
$z$-axis such that its rotational velocity in the $x$-$y$ plane is 
$v_\phi = v_{\rm rot} = f_{\rm rot} \sqrt{2} \sigma$ with $f_{\rm rot}=0.3$; it falls
from rest in the radial direction. Finally we 
embed a collisionless particle -- corresponding to the black hole -- at rest 
at the centre of the potential; the initial black hole mass is 
$M_{\rm BH}=10^6\,\rm M_{\odot}$ and, in the cases where we use the accretion 
disc particle model, an initially zero disc mass.

All of the simulations are run using {\small GADGET3}, an updated version 
of the code presented in \citet{springel.2005}. Each simulation is run for 
$\sim$ 4.7 Myrs, which corresponds to $\sim$ 14 dynamical times at the initial 
outer radius of the shell.

\subsection{Without Feedback}

\begin{figure}
  \centerline{\psfig{file=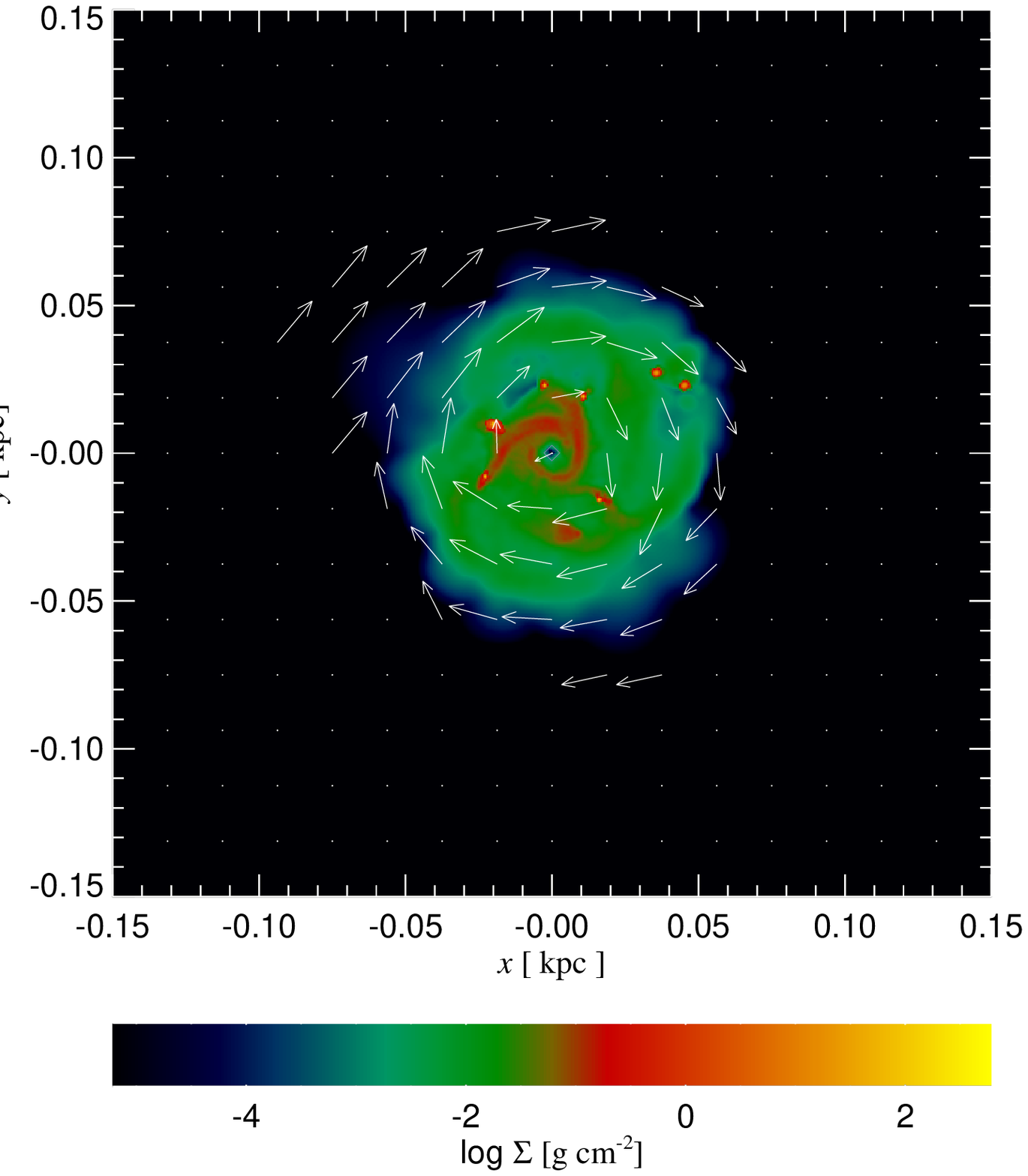,width=0.4\textwidth,angle=0}}
  \centerline{\psfig{file=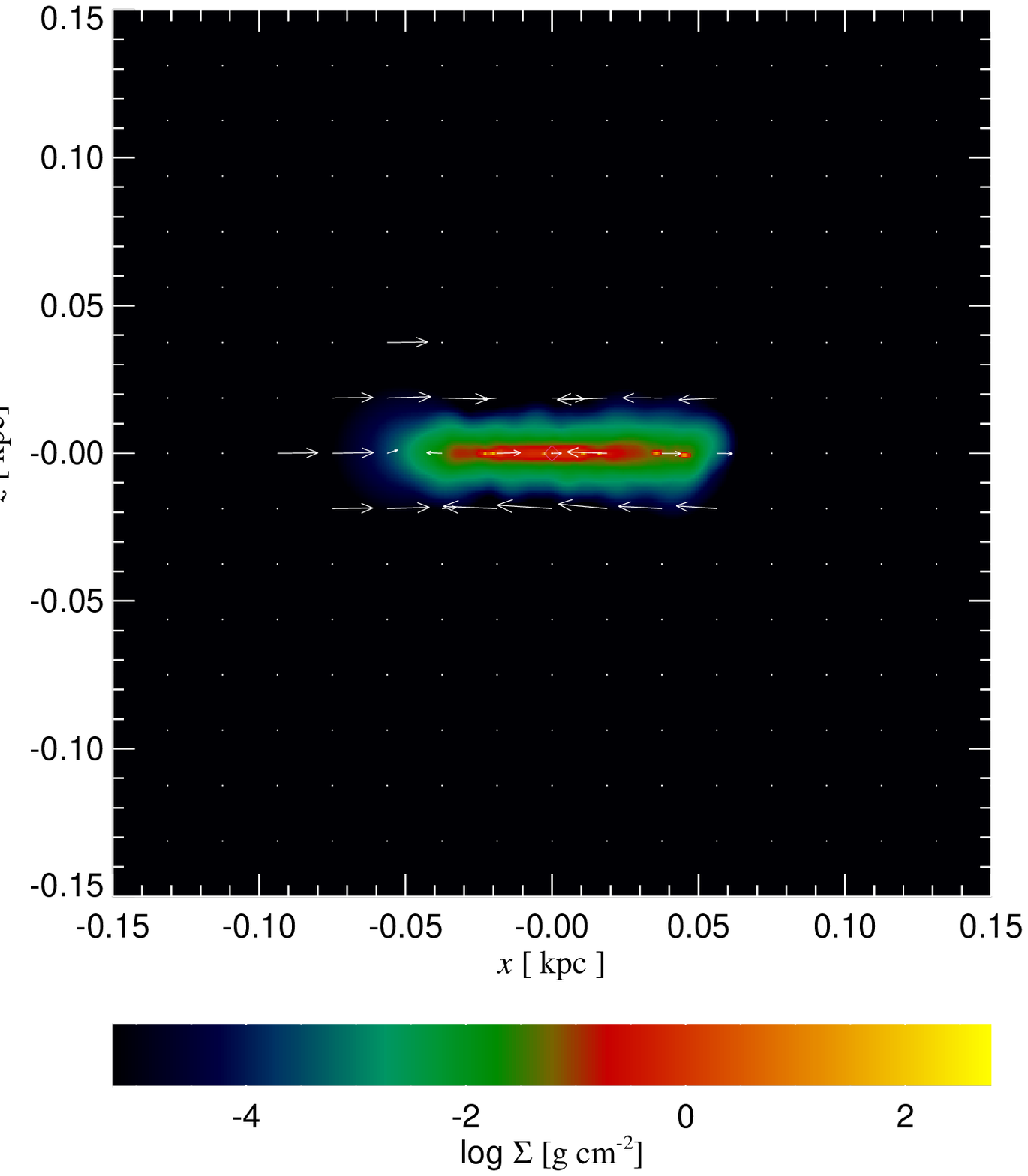,width=0.4\textwidth,angle=0}}
  \caption{{\bf Without Feedback :} The projected gas density in the $x$-$y$ 
    and $x$-$z$ planes at $t$=4.7 Myrs (upper and lower panels respectively). 
    As before, arrows indicate the velocity vectors of the gas. Note 
    the dense knots corresponding to high density regions of sink 
    particle formation. The ADP has an accretion radius of $R_{\rm acc}=0.003$ 
    kpc.}
\label{fig:no_feedback}
\end{figure}

We begin by considering the simplest possible case -- the collapse of the
shell in the absence of any feedback from the black hole. We model the black
hole using an ADP, but in this particular
simulation we decouple the accretion luminosity $L_{\rm acc}$ of the black
hole from $\dot M_{\rm BH}$ by setting $\eta=0$ in equation~\ref{eq:luminosity};
this suppresses both the momentum-driven wind and quasar pre-heating. We 
choose an accretion radius of $R_{\rm acc}=0.003\,\rm kpc$\footnote{This is 
larger than the gravitational softening of gas particles by a factor of 
$\sim 20$, but slightly smaller than the inner edge of large-scale gas disc 
we expect to form. We choose a large value to highlight that our initial 
conditions minimal accretion onto the black hole, by construction.}. For 
simplicity, we assume an isothermal equation of state with a temperature of 
$T=10^4$ K.

By conservation of angular momentum, the shell should settle into a thin
rotationally supported disc \citep[cf.][]{hobbs.etal.2010a}. This disc is
distinct from, and on a much larger scale than, the accretion disc discussed in
\S\ref{ssec:accretion}, which is tightly bound to the black hole on a scale 
much smaller than we can resolve in our simulation. We show the gas density 
projected onto the $x$-$y$ and $x$-$z$ 
planes at $t$=4.7 Myrs (upper and lower panels respectively) in 
Fig~\ref{fig:no_feedback}. The gas is distributed in a thin 
rotating disc; its inner and outer boundaries are at $\sim 0.006$ and 
$\sim 0.01$ kpc respectively and it rotates in a clock-wise sense around the 
$z$-axis (indicated by the projected velocity vectors). The inner boundary is 
larger than the accretion radius $R_{\rm acc}=0.003\,\rm kpc$ by a factor of 
$\sim$2 and so only a small fraction of the mass of the disc comes within 
$R_{\rm acc}$ over the duration of the simulation ($\sim$20 particles or 
$\sim\,0.007\%$ after $\sim\,4$ Myrs). 

The absence of any accretion until late times might seem counterintuitive, 
given that our initial condition is a rotating shell of gas. What happens
to material with small angular momentum that lies along the axis of rotation?
Why does it not accrete rapidly onto the ADP? \citet{hobbs.etal.2010a} have 
shown that this small angular momentum gas shocks and mixes with larger angular
momentum gas, which increases its net angular momentum and provides a barrier 
to accretion.

Note that there are knots of high density material in the disc. For 
expediency, we tag gas particles that exceed a threshold value in local 
density. Physically these high density regions are likely to host star 
formation, but for the purpose of this study we simply decouple these 
particles hydrodynamically from other gas particles, ignoring them in the 
hydrodynamical force calculation, which helps to increase the speed of 
the simulation. In this particular run, $\sim 90\%$ of the gas particles are
converted to decoupled particles by $t \simeq$ 4.7 Myrs.

\subsection{With Feedback}

\begin{figure*}
  \centerline{
    \psfig{file=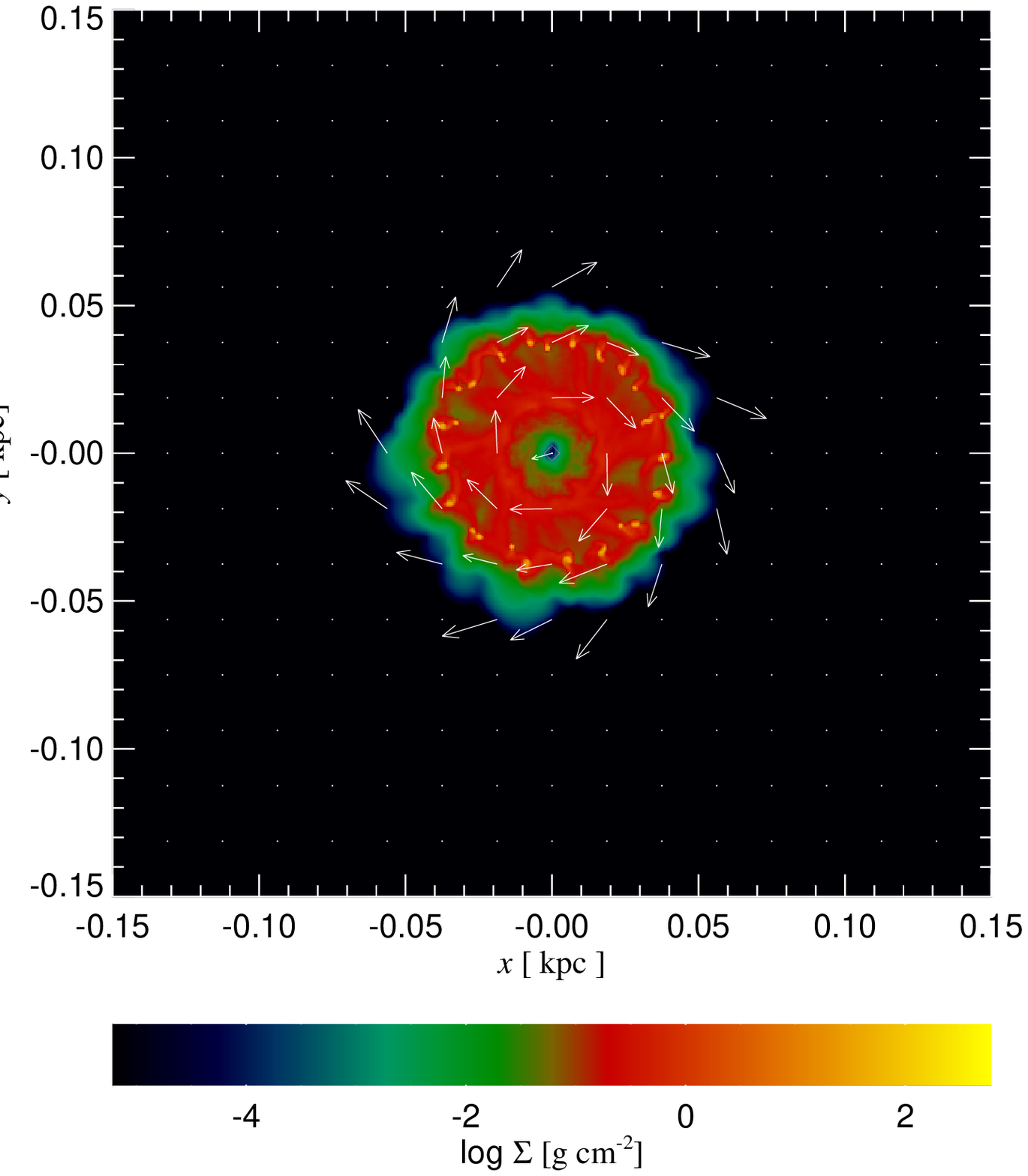,width=0.47\textwidth,angle=0}
    \psfig{file=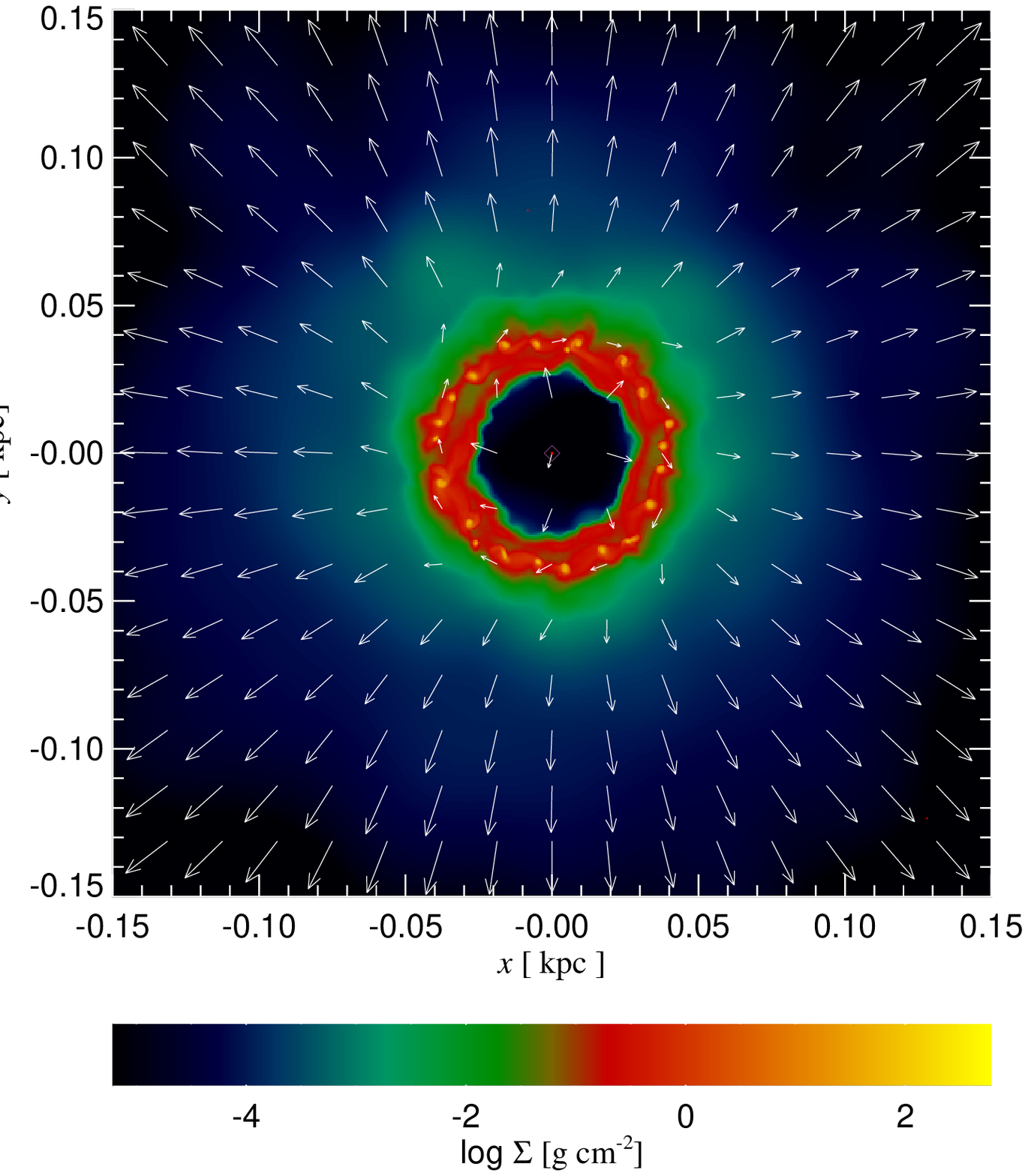,width=0.47\textwidth,angle=0}}
  \centerline{
    \psfig{file=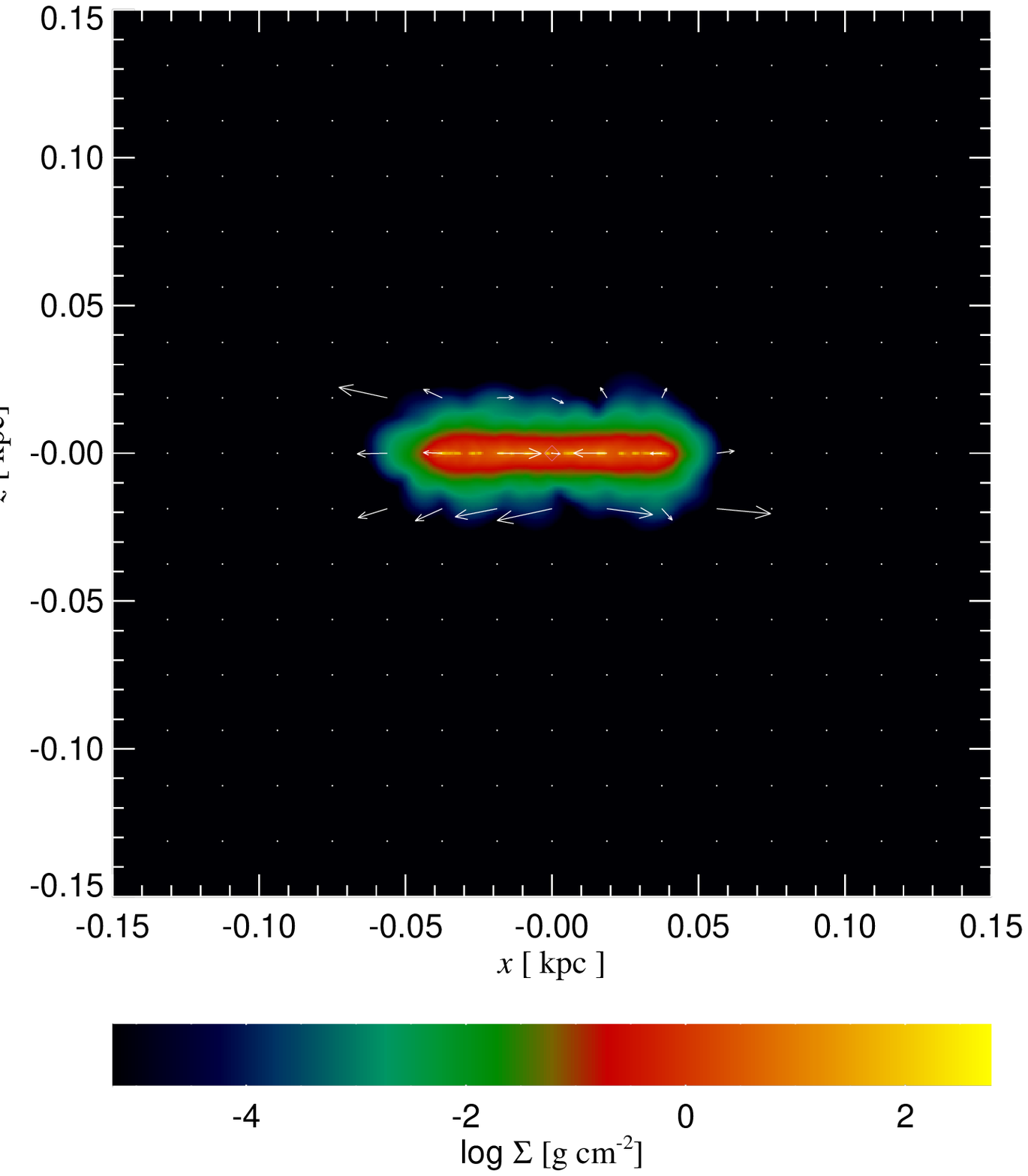,width=0.47\textwidth,angle=0}
    \psfig{file=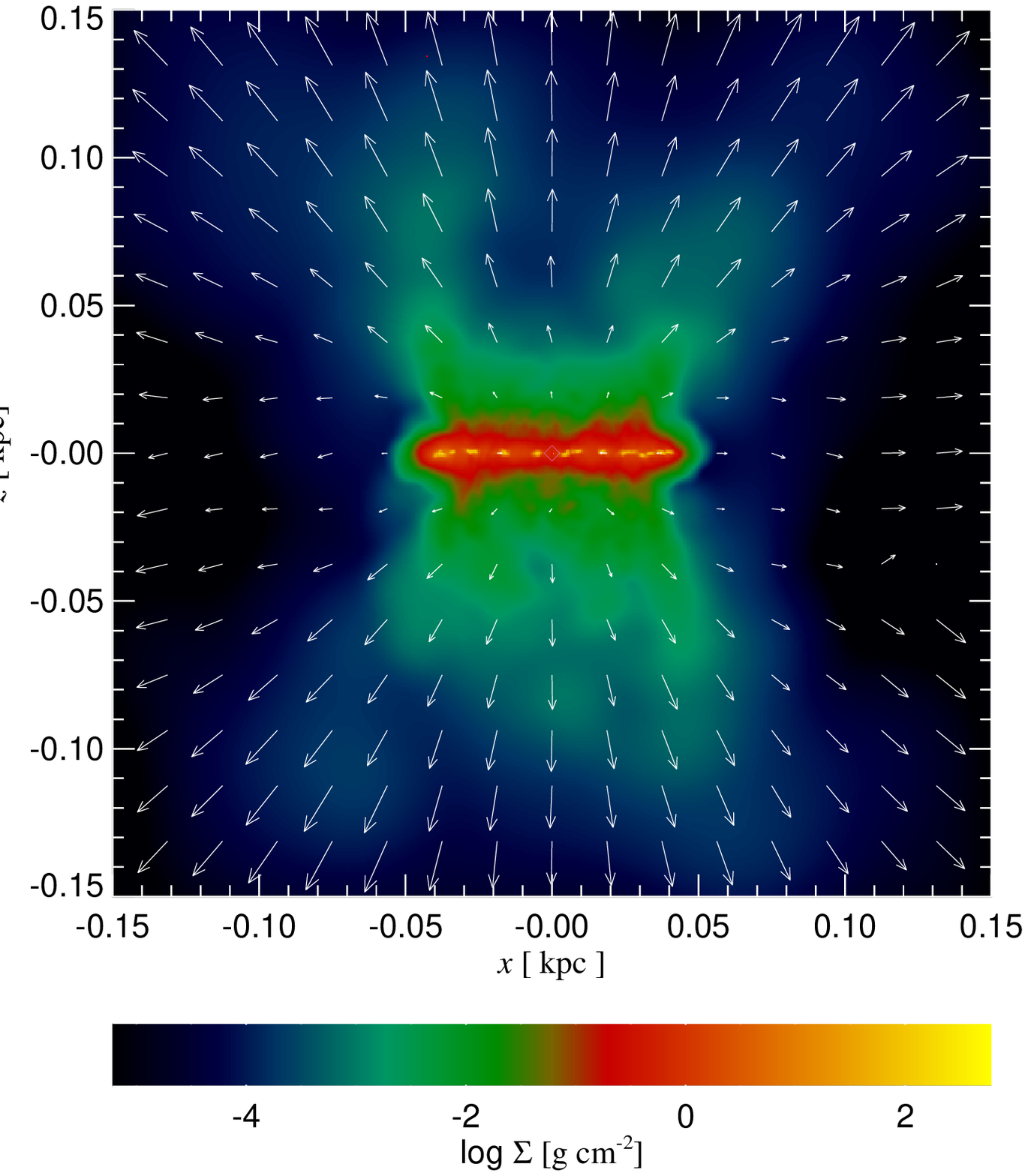,width=0.47\textwidth,angle=0}}
  \caption{{\bf Early Times:} The gas density projected onto the $x$-$y$ and 
    $x$-$z$ planes (upper and lower panels) in the accretion disc particle and
    Bondi-Hoyle runs (left and right panels) at $t \simeq 1$ Myr. }
\label{fig:early_times}
\end{figure*}

\begin{figure*}
  \centerline{
    \psfig{file=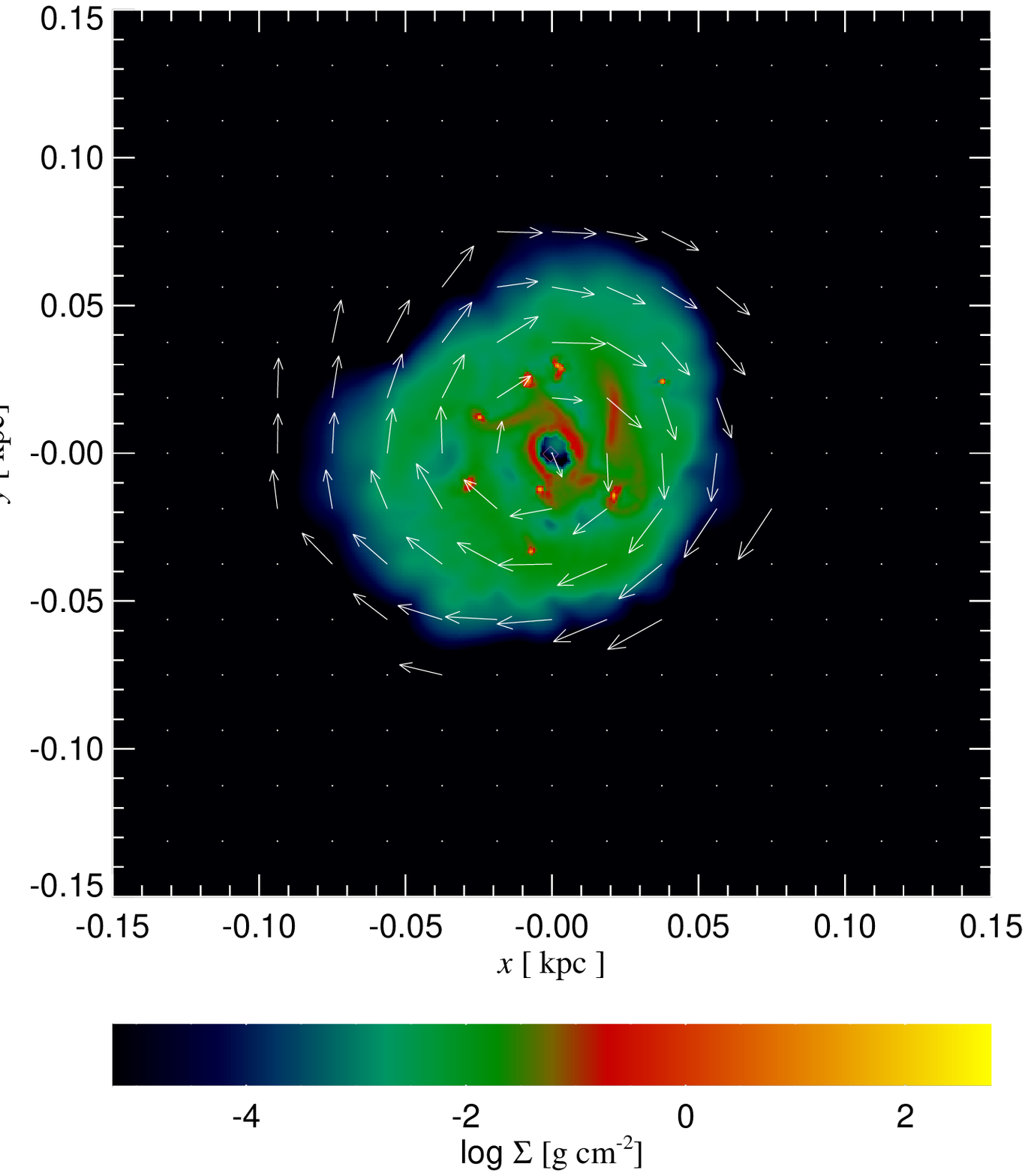,width=0.47\textwidth,angle=0}
    \psfig{file=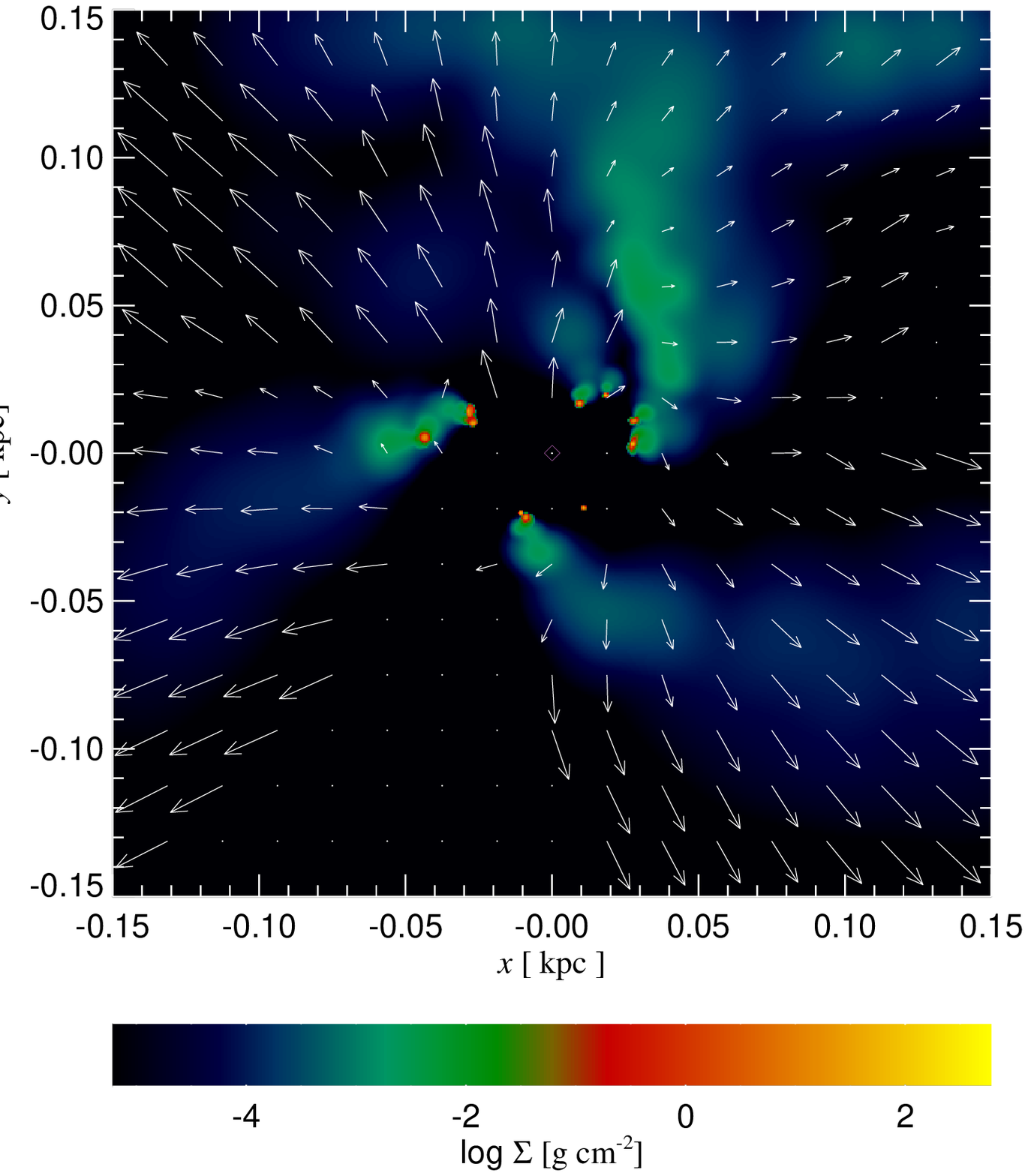,width=0.47\textwidth,angle=0}}
  \centerline{
    \psfig{file=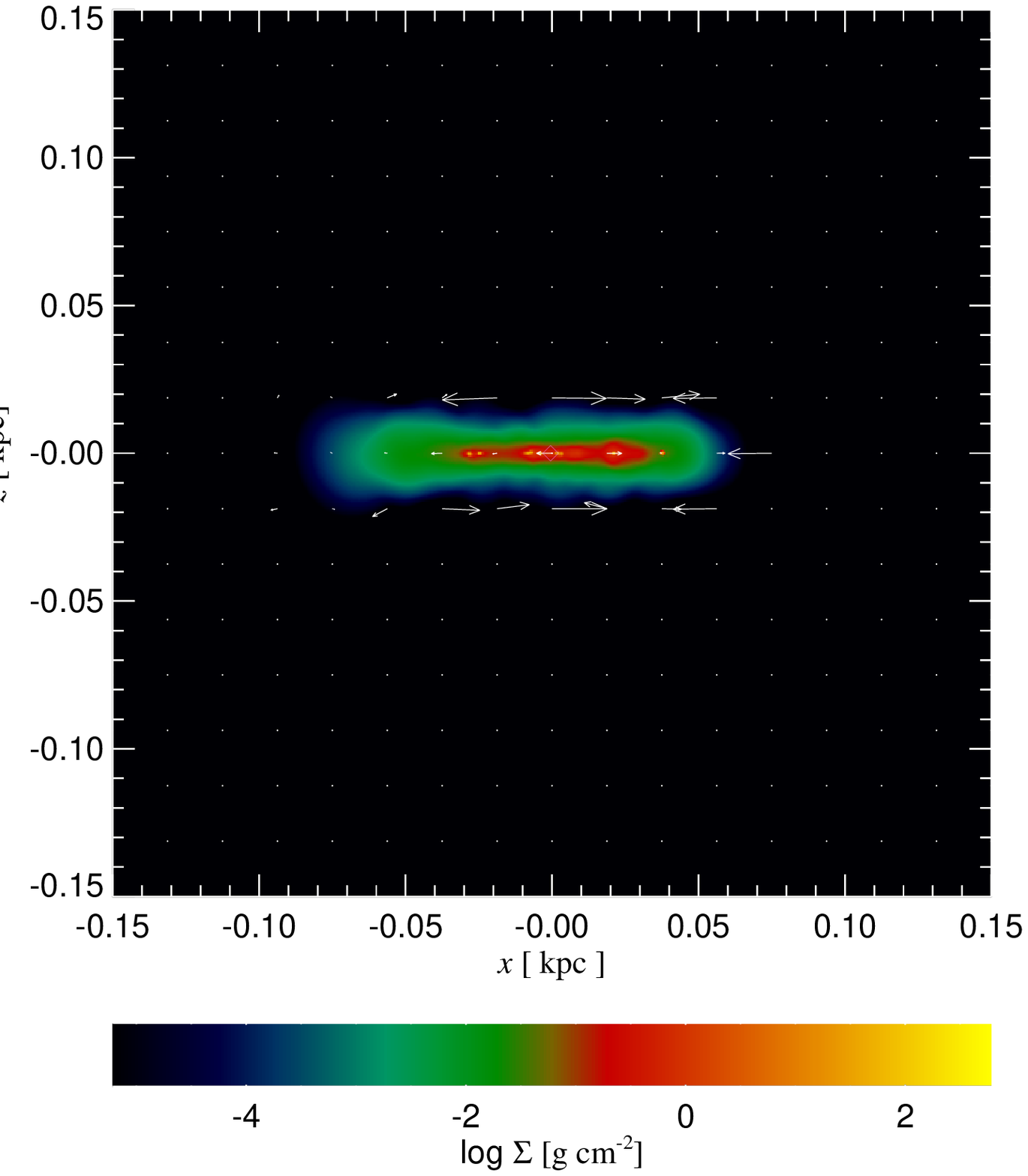,width=0.47\textwidth,angle=0}
    \psfig{file=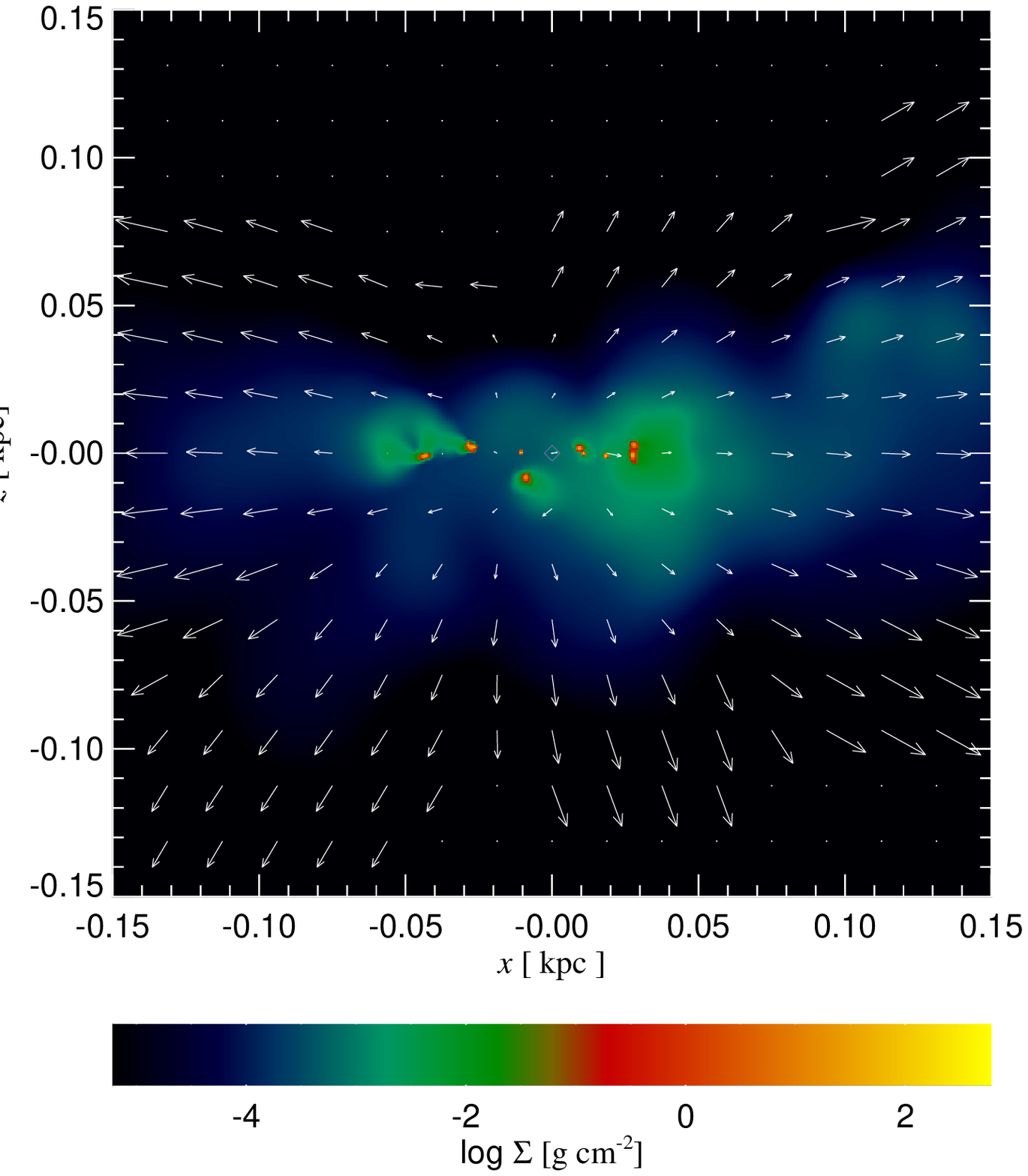,width=0.47\textwidth,angle=0}}
  \caption{{\bf Late Times:} The gas density projected onto the $x$-$y$ and 
    $x$-$z$ planes (upper and lower panels) in the accretion disc particle and
    Bondi-Hoyle runs (left and right panels) at $t \simeq 4.7$ Myrs. }
\label{fig:late_times}
\end{figure*}

Let us now consider the evolution of the collapsing shell when the black hole
accretion luminosity $L_{\rm acc}$ is coupled to $\dot M_{\rm BH}$, as in 
equation~\ref{eq:luminosity}. We estimate $\dot M_{\rm BH}$ using the 
ADP method (equations \ref{eq:dmdiscdt} and 
\ref{eq:dmbhdt} with $R_{\rm acc}=0.003$ kpc and $t_{\rm visc} \simeq 10^4$ yrs) 
and the Bondi-Hoyle method (equation~\ref{eq:bondi_hoyle} with $\alpha$=1). 
Note that, for the purpose of this study, the precise value of $t_{\rm visc}$ 
is unimportant; the point is that in the cases that we consider, the ADP 
method predicts a negligible accretion rate, as we would expect 
from physical arguments. In both cases we assume that the black hole feedback 
takes the form of a momentum-driven wind and quasar pre-heating. The gas is
initially isothermal with a temperature of $T=10^4$ K, but as it evolves it
can heat and cool in response to, for example, the quasar radiation field.

In Fig.~\ref{fig:early_times} and Fig.~\ref{fig:late_times} we show the 
gas density projected onto the $x$-$y$ and $x$-$z$ planes (upper and lower 
panels respectively) in the ADP and Bondi-Hoyle runs 
(left and right panels) at $t$=1 Myr and $t$=4.7 Myrs. As in 
Fig.~\ref{fig:no_feedback}, arrows indicate the magnitude and direction 
of the projected velocity vectors of the gas. The differences between the 
runs are striking. The shell should settle into a thin 
rotationally supported disc whose properties are very similar to those of 
the disc shown in Fig.~\ref{fig:no_feedback} and indeed this is the case 
in the ADP run. This is unsurprising -- because we link 
feedback explicitly to accretion rate onto the black hole, we do not expect 
any significant feedback in the ADP run because the 
angular momentum of the gas is too large to bring it within $R_{\rm acc}$ until 
late times. Even at this point, the mass of gas accreted $M_{\rm acc}\ll 1\%$
over the lifetime of the simulation and $\dot M_{\rm BH} \ll \dot M_{\rm Edd}$, 
which means that the feedback is weak and has little effect on the gas 
distribution. We note also that, as in the run without feedback, $\sim 92\%$ 
of the gas particles have decoupled into high density knots by $t\simeq 
4.7$ Myrs.

\begin{figure}
  \centerline{\psfig{file=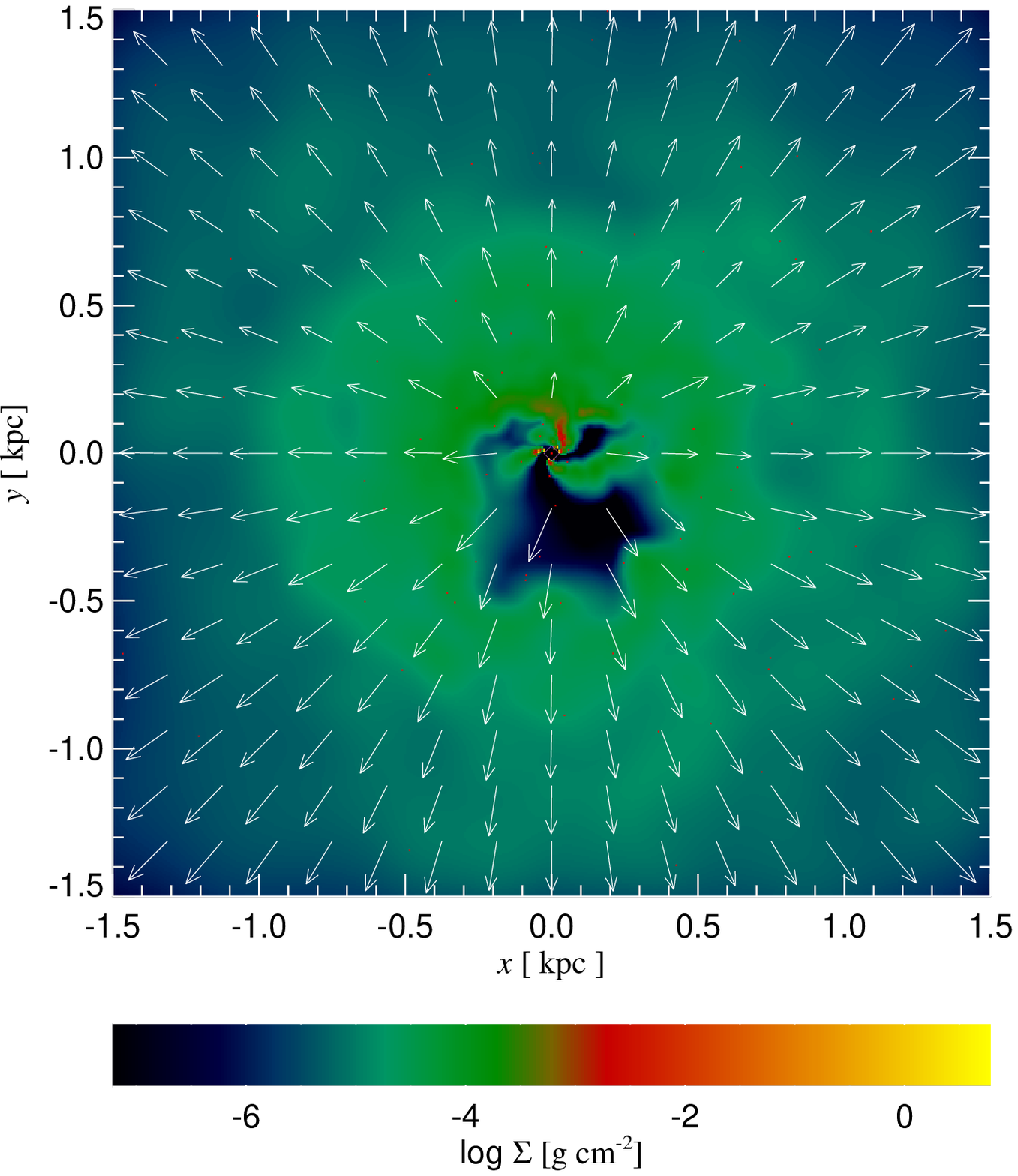,width=0.47\textwidth,angle=0}}
  \centerline{\psfig{file=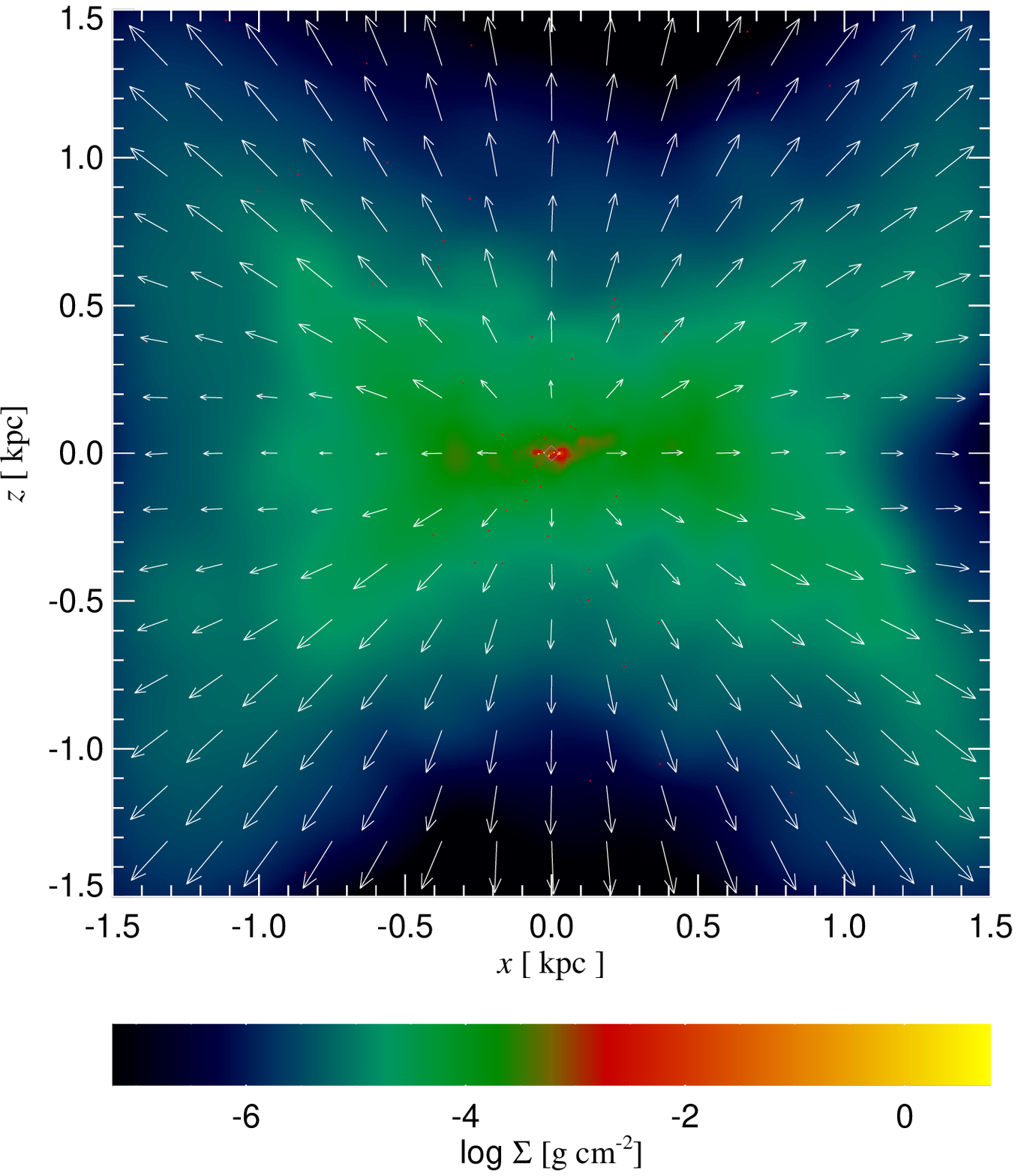,width=0.47\textwidth,angle=0}}
  \caption{Large-scale distribution of gas in the Bondi-Hoyle run at $t \simeq 
    4.7$ Myrs.}
\label{fig:bondi_hoyle_largescale}
\end{figure}

In contrast, the accretion rate is consistently Eddington limited over the 
duration of the simulation in the Bondi-Hoyle run. The black hole is not massive
enough for its feedback to prevent the collapse of the shell into a disc
(cf. right hand panels of Fig.~\ref{fig:early_times}), but once the disc 
has formed, the feedback acts efficiently on the low column density gas.
At early times it is the lower column density gas surrounding the disc and 
along the axis of rotation that is most efficiently driven outwards, 
principally by the momentum-driven wind. In particular, it is the 
impact of the feedback along the axis of rotation that imprints the strongly 
bipolar character on the outflow, evident in the $x-z$ projection in
Fig.~\ref{fig:early_times}. Over time, as high density knots form 
in the disc, lower density material within the disc is blown away. For example,
after $t \simeq$ 1/2.8/4.7 Myrs, $\sim$2\%/25\%/40\% of the gas particles that 
have not been decoupled have been expelled from the disc, compared to 
$\ll 1\%$ of gas particles in the accretion disc particle run. At late times,
even the high density knots are ablated. As a result, the cumulative 
effect of the feedback over $\sim 4.7$ Myrs has a profound impact on the gas 
distribution, a point that is nicely illustrated in 
Fig.~\ref{fig:bondi_hoyle_largescale}, which makes clear that the gas is 
distributed over kpc-scales (and indeed to $\sim 10$ kpc) by the end 
of the run. 

It is worth noting that the fraction of gas that had a sufficiently high 
density to be decoupled over the lifetime of the Bondi-Hoyle run is 
comparable to the fraction in the accretion disc particle run ($88\%$ 
compared to $90\%$). However, these high density knots are ablated by the 
feedback in the Bondi-Hoyle run whereas they survive in the accretion disc 
particle run.

\section{Summary}
\label{sec:summary}

Black holes grow by accreting gas and stars from their surroundings. However,
only the lowest angular momentum material can come sufficiently close to the 
black hole to be accreted, and so any estimate of a black hole's accretion rate 
$\dot M_{\rm BH}$ must account for this. However, the standard sub-grid model
for black hole accretion in galaxy formation simulations neglects the 
angular momentum of accreting material \citep[cf.][]{dimatteo.2005,
springel.etal.2005}. The Bondi-Hoyle method \citep{bondi.hoyle.1944,bondi.1952}
assumes that $\dot M_{\rm BH} \propto \rho/c_s^3$ where $\rho$ is the gas 
density at the position of the black hole and $c_s$ is the sound speed in the 
gas. This implies that black holes are always accreting; $\dot M_{\rm BH}$ may 
be small but it can never be zero, regardless of the angular momentum of the 
gas surrounding the black hole.

In this short paper, we presented a new sub-grid model for estimating 
$\dot M_{\rm BH}$  in galaxy formation simulations that accounts for the 
angular momentum of accreting material. This accretion disc particle (ADP) model
uses a collisionless sink particle to model the composite black hole and 
accretion disc system. The black hole accretes if and only if gas comes within 
the accretion radius $R_{\rm acc}$ of the ADO, at which point its mass is added 
to the accretion disc that feeds the black hole on a viscous timescale 
$t_{\rm visc}$. In this way the black hole will accrete only the lowest angular 
momentum material available to it in and around its host galaxy. 

We demonstrated that the ADP method constitutes a physically self-consistent 
model using simple idealised numerical simulations that follow the collapse 
of a rotating shell of gas onto a black hole embedded at the centre of an 
isothermal galactic potential. By construction, the gas settles into a thin 
rotationally supported disc between $R_{\rm in}$ and 
$R_{\rm out}$, where $R_{\rm min} > R_{\rm acc}$, and so we do not expect any 
accretion onto the black hole. Because we link feedback to accretion, we do
not expect there to be any significant differences between simulations with or
without feedback when using the ADP estimate of 
$\dot M_{\rm BH}$. On the other hand, we expect the evolution of the system to 
differ if $\dot M_{\rm BH}$ is estimated using the Bondi-Hoyle method.

These expectations were borne out by the results of our simulations. The 
Bondi-Hoyle method predicted that $\dot M_{\rm BH}$ should 
be Eddington limited over the lifetime of the simulation. Because the feedback 
in this case was relatively weak, its effect could not prevent the collapse of 
the shell into a disc, but the cumulative effect of the feedback was to drive
gas away and to expel it to $\sim \rm 10 kpc$ scales after $\sim 5$ Myrs. In 
contrast the ADP method predicted negligible accretion 
rates at all times; the shell collapsed, settled into a thin rotationally 
supported disc and $\sim 90\%$ of the mass decouples into long-lived high 
density knots, which correspond to regions of star formation.

\section{Conclusions}
\label{sec:conclusions}

We have argued that our new accretion disc particle (ADP) method provides a far 
more physically motivated and self-consistent approach to modelling black hole
accretion than the Bondi-Hoyle method, which is the standard approach in 
galaxy formation simulations 
\citep[cf.][]{springel.etal.2005,dimatteo.2005}. The Bondi-Hoyle method 
was formulated with a specific astrophysical problem in mind, quite unlike 
the situations that arise when modelling galaxy formation. It is not 
applicable to problems 
in which the accretion flow has non-zero angular momentum (as demonstrated 
in this paper) and/or in which it is embedded in the potential of a more 
massive host (as we show in Hobbs et al., in preparation). Therefore it is 
unsurprising that the Bondi-Hoyle method struggles to capture the 
behaviour of gas accretion in these kinds of common situations.
Our ADP method is similar in spirit to the 
``accretion radius'' or ``sink particle'' approaches to modelling
accretion that are used extensively in simulating star formation 
\citep[e.g.,][]{bate.1995,bate.bonnel.2005} and modelling gas accretion 
onto the super-massive black hole at the centre of the Milky Way
\citep{cuadra.etal.2006}, and we believe that it is natural to extend this
approach into modelling galaxy formation. An important next step in our 
work, which builds on this accretion disc particle method and our recent 
momentum-driven wind model for feedback \citep[cf.][]{nayakshin.power.2010}, 
is to combine the models in simulations of merging galaxies and ultimately 
cosmological galaxy formation simulations.\\

It is interesting to consider one important astrophysical consequence of our
accretion disc particle model and to contrast it with what one would expect
using the Bondi-Hoyle model. Recently it has been suggested that there is
observationally evidence for separate $M_{\rm BH}-\sigma$ relations for 
elliptical galaxies and classical bulges on the one hand and pseudo-bulges 
on the other, such that the black holes in pseudo-bulges are underweight
\citep[see, for example,][]{greene.2008,hu.2009}. The properties of 
pseudo-bulges appear to deviate systematically from those of classical 
bulges, and in particular they are characterised by a high degree of rotational
support. As we have shown, the angular momentum of infalling material
provides a natural barrier to black hole growth, and so we would expect that
rotationally supported systems to be more likely to be 
systems in which the central super-massive black hole is malnourished 
and underweight. Of course, the precise details of a galaxy's assembly 
history are important but our model would predict a systematic offset between
super-massive black hole masses in galaxies that have on average accreted 
higher angular momentum material than those that have on average accreted lower
angular momentum material. In contrast, the Bondi-Hoyle model would predict 
that the black hole should continue to grow to a critical black hole mass 
imposed by the depth of the gravitational potential in which it sits, regardless
of the angular momentum of infalling material. We shall investigate this
question further in future work.

\section{Acknowledgments}

Theoretical astrophysics research at the University of Leicester is supported 
by a STFC Rolling grant.

\bibliographystyle{mnras}
\bibliography{references}

\label{lastpage}

\end{document}